\documentclass{egpubl}
\usepackage{pg2026s}

\WsConferencePaper

\usepackage[T1]{fontenc}
\usepackage{dfadobe}  

\biberVersion
\BibtexOrBiblatex
\usepackage[backend=biber,bibstyle=EG,citestyle=alphabetic,backref=true]{biblatex}
\electronicVersion
\PrintedOrElectronic
\ifpdf \usepackage[pdftex]{graphicx} \pdfcompresslevel=9
\else \usepackage[dvips]{graphicx} \fi

\usepackage{egweblnk}

\usepackage{amsmath}
\usepackage{amssymb}
\usepackage{tikz}
\usepackage{caption}
\usepackage[percent]{overpic}
\usepackage{makecell}
\usepackage{booktabs}
\usepackage{multirow}

\newcommand{\rev}[1]{#1}

\title[CT2Yarn: Yarn-Level Reconstruction of Crochet from Computed Tomography]%
      {CT2Yarn: Yarn-Level Reconstruction of Crochet \\from Computed Tomography}

\author[Chang Luo \& Nobuyuki Umetani]
{\parbox{\textwidth}{\centering Chang Luo\orcid{0009-0006-2717-4598}
        and Nobuyuki Umetani\orcid{0000-0003-1251-970X}
        }
        \\
{\parbox{\textwidth}{\centering The University of Tokyo, Japan
       }
}
}

\begin{document}

\teaser{
 \centering
 \includegraphics[width=\linewidth]{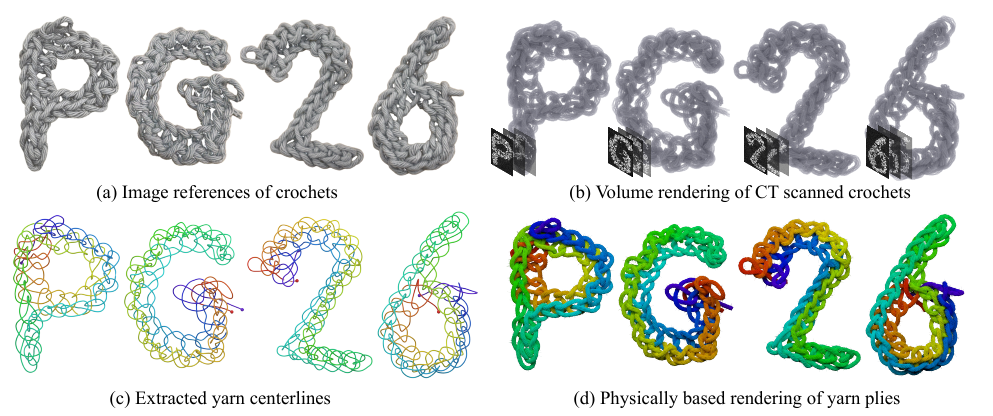}
 \caption{\textbf{From real crocheted samples to a continuous yarn path.}  Starting from real crocheted samples (a), we acquire micro-CT scans that reveal their internal structure (b).  Our pipeline then reconstructs a single continuous yarn centerline (c), visualized with colors indicating arc length, and generates a ply-level yarn geometry for rendering (d).}
\label{fig:teaser}
}

\maketitle
\begin{abstract}
We introduce CT2Yarn, a \rev{human-in-the-loop} framework for recovering a single continuous yarn path from micro-computed tomography (micro-CT) scans of real crochet objects.
Crochet is a craft that creates complex three-dimensional shapes by interlocking loops formed from a single yarn. 
Recovering the underlying yarn path from external observations is challenging because of severe self-occlusion. While micro-CT reveals the full internal structure of a crochet object, the volumetric scan alone does not explicitly encode how the yarn traverses the object.
This challenge stems from the hierarchical structure of yarn: a yarn consists of multiple twisted plies, and each ply itself consists of twisted fibers. Consequently, local fiber orientations observed in micro-CT scans are not aligned with the overall yarn direction.
To recover yarn-level orientations from ply-level fiber orientations, we first estimate local fiber directions using Gabor filtering and convert the volume into an oriented point cloud. 
We then introduce an anisotropic mean-shift procedure that aggregates local fiber orientations within a neighborhood of the yarn radius into yarn-level orientation estimates.
Combined with an automatic topology skeletonization strategy, our method extracts yarn-path fragments.
\rev{Subsequent fragment linking, junction cleaning, and loop detection merge these trees into a small number of long curves.}
\rev{Where the automatic reconstruction remains ambiguous, a sketch-based user interface enables users to interactively complete the single continuous yarn path.}
The recovered yarn path enables downstream applications including physically based simulation, ply-level rendering, and stitch-pattern extraction.
\rev{Our source code is publicly available at \url{https://github.com/netbeifeng/ct2yarn}.}


\end{abstract}  
\section{Introduction}
\label{sec:intro}
Crochet is a textile craft that creates complex three-dimensional structures by interlocking loops formed from a single continuous yarn.
Owing to its flexibility and accessibility, crochet is widely used to produce garments, plush toys, and artistic works.
In computer graphics, various studies have explored the physically based simulation~\cite{knit_sim_yarn_08,
knit_stitchmesh_orig_12} and
photorealistic rendering~\cite{cloth_render_surface_23, knit_sim_render_rt_25} using explicit yarn-path representations.
However, almost all computational work on yarn modeling has focused on synthetic structures~\cite{knit_stitchmesh_fab_19,
crochet_stitchmesh_20, crochet_amigo_22}, where the yarn path is procedurally generated from an idealized stitch pattern.
Recovering the yarn path of real crocheted objects is essential for faithfully digitizing handmade crochet and enabling downstream applications such as simulation and rendering, yet remains largely unexplored.

A natural starting point for recovering yarn structure from real samples is image-based reconstruction.
Recent work has made progress on woven
fabrics~\cite{woven_capture_photo_22, woven_capture_fiber_26} and on knit fabrics~\cite{knit_inverse_neural_19}, but all assume the fabric forms a regular repeating grid.
However, unlike knitting, where stitches are connected in a regular row-by-row manner, crochet stitches can be worked into arbitrary earlier stitches, producing freeform topologies that cannot be naturally represented by a regular grid.
Moreover, interlocked loops occlude a substantial portion of the yarn from every viewpoint.
Consistent with these challenges, existing crochet-specific tracking systems rely on hand-worn motion sensors rather than visual
inputs~\cite{crochet_track_motion_25}, and recent benchmarking efforts indicate that vision-language models fail to reliably recover abstract crochet patterns from images~\cite{crochet_bench_vlm_25}.

Micro-computed tomography (micro-CT) avoids the limitation of occlusion by directly imaging the interior of a physical sample using X-rays.
Previous work has used this volumetric imaging modality to reconstruct woven
fabric~\cite{woven_ct_appearance_11, woven_ct_proc_16} and
hair~\cite{hair_ct_recon_23}.
Although modern CT scanners provide voxel resolutions well below typical yarn diameters, recovering yarn structure from the resulting density volumes remains challenging.
Unlike hair, whose strands are separated in air with clear boundaries, neighboring yarn segments in crochet are brought into tight contact at every stitch, causing their boundaries to become ambiguous (Fig.~\ref{fig:yarn_vs_hair}).
Furthermore, the fiber orientations within a yarn are not aligned with the yarn direction itself. 
A yarn consists of multiple twisted plies, each of which is in turn composed of twisted fibers.
Consequently, the local fiber orientations observed in the CT volume provide only indirect information about the underlying yarn trajectory.
Together, these challenges make individual yarn paths difficult to distinguish in the volume. 
Furthermore, crocheted objects exhibit a strong topological constraint: they are constructed from a single continuous yarn.
Unlike hair reconstruction, where strands are largely independent, an erroneous connection in crochet can propagate globally and corrupt the recovered yarn path.

\begin{figure}[t]
  \centering
  \includegraphics[width=\linewidth]{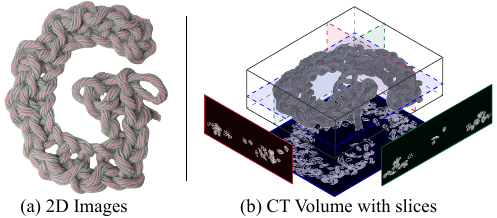}
  \caption{\textbf{Photograph and micro-CT volume of the same
  crocheted sample.} (a) Interlocking loops occlude most of the
  interior yarn from any single viewpoint. (b) The CT volume,
  rendered with three orthogonal slices, exposes every yarn
  cross-section throughout the fabric.}
  \label{fig:ct_input}
\end{figure}

\rev{We present \textit{CT2Yarn}, a human-in-the-loop framework that combines an automatic reconstruction pipeline with lightweight interactive editing to recover a single continuous yarn path from the micro-CT density volume of a crocheted object.}
Given a 3D density volume as input, our method reconstructs a centerline curve representing the entire yarn.
To account for the hierarchical structure of yarn, which consists of multiple plies, our yarn orientation estimation proceeds in two stages.  
We first estimate a voxel-wise orientation field of local plies using Gabor filtering.
We then infer yarn-level orientations from the estimated ply orientations using an anisotropic mean-shift clustering procedure.
The yarn centerline candidates are extracted by connecting neighboring mean-shift points into a graph.
However, the recovered centerline candidates are fragmented due to the tight contacts between neighboring yarn segments. 
We therefore resolve this by a heuristic linking startegy and perform cleaning as the automatic pipeline.
The automatic result provides a good starting point, and an interactive interface lets users resolve the remaining ambiguous connections, completing the final single continuous yarn path.

We evaluate our method on micro-CT scans of physical crocheted samples
covering various stitch types.
\rev{We evaluate primarily through visual comparison, complemented by a quantitative measures, ablation studies of the main algorithmic components, and a demonstration of stitch-pattern extraction via template fitting. 
As no ground-truth yarn geometry is available for real crocheted objects, the quantitative measures are computed against manually annotated references, and a synthetic CT volume with an exact ground truth. 
We also quantify the manual effort of the interactive completion, at a median of four sketch edits and under three minutes per sample.}
The resulting yarn representation supports downstream applications including physically based simulation, ply-level rendering, and stitch-pattern extraction.
To support future research, we release both the micro-CT volumes and the reconstructed yarn centerline curves.
Our contributions are as follows:
\begin{itemize}
  \item A two-stage orientation analysis that infers yarn-level trajectories from ply-level fiber orientations observed in micro-CT volumes.
  \item \rev{An automatic repair stage that merges fragmented centerline candidates into a small number of long curves, and a sketch-based interface for the few connections it cannot resolve.}
\item A publicly available dataset consisting of micro-CT scans and reconstructed yarn curves of real crocheted samples.
\end{itemize}

\section{Related Work}
\label{sec:related_work}

\subsection{Yarn-Level Modeling of Knitted Fabrics}
\label{subsec:knit}
The regular row-and-column structure of knit stitches has enabled an extensive yarn-level pipeline in computer graphics. The stitch mesh~\cite{knit_stitchmesh_orig_12} represents knit
fabric as a quad-dominant mesh whose faces correspond to stitches, and has been extended to automatic generation from arbitrary 3D shapes~\cite{knit_stitchmesh_auto_18} and to knittability-preserving editing~\cite{knit_stitchmesh_fab_19}. 
Industrial knitting machines are themselves driven by formal compilers and visual programming interfaces~\cite{knit_machine_compiler_16, knit_machine_visual_19},
enabling automated fabrication of complex 3D knit garments.
\rev{Solid knitting extends this automated fabrication from hollow garments to dense volumetric objects, and the interactive design tool skCAD~\cite{hirose:2026:skcad} lets users design such solids on a three-dimensional stitch lattice with automatically generated machine-executable patterns.}

Yarn-level simulation, initiated by Kaldor et al.~\cite{knit_sim_yarn_08}, has been refined through persistent contact handling~\cite{knit_sim_contact_17}, accelerated via homogenization~\cite{knit_sim_homogenize_20, knit_sim_homogenize_24}, and embedded in interactive tools for authoring and relaxing knit and woven patterns on a periodic grid~\cite{knit_sim_interactive_18}. 
Alongside simulation, fiber- and ply-level rendering models reproduce
knit and woven appearance at varying scales~\cite{cloth_render_fiber_rt_19,
cloth_render_surface_23, knit_sim_render_rt_25}, with recent neural
approaches further accelerating woven fabric rendering~\cite{woven_render_neural_24}.

This mature pipeline is built around a single structural assumption: that the fabric is organized as a regular grid. 
It lets every stitch be addressed by row-and-column indices, a rule shared by all of the pipelines above.
Crochet violates this assumption: its stitches can be worked into any earlier stitch, producing freeform topologies that no strict row-and-column grid captures.

\subsection{Computational Approaches to Crochet}
\label{subsec:crochet}

Computational work on crochet is more recent and smaller in scope, focusing on forward synthesis, pattern representations, machine fabrication, and process tracking.
Early efforts convert target geometric forms into crocheting recipes~\cite{crochet_stitches_geom_17}, adapt the knit stitch mesh representation to produce yarn-level geometry~\cite{crochet_stitchmesh_20}, and map a 3D character mesh to a crochetable pattern for amigurumi~\cite{crochet_amigo_22}, following the interactive plush-design tool Knitty~\cite{igarashi2008knitty}. 
Storck et al.~\cite{crochet_topology_22} model each stitch as a unit cell of parameterized key points along the yarn centerline, interpolated by Kochanek-Bartels splines and exported to finite element simulation. 
Beyond synthesis, Seitz et al.~\cite{crochet_pattern_dsl_22} introduce a visual domain-specific language for patterns, ~\cite{crochet_machine_designtool_23, crochet_machine_pattern_23} develop a design tool for the crochet machine prototype, StitchFlow~\cite{crochet_track_motion_25} infers the crochet progress from inertial measurements of the crafter's hand.

None of these works recovers the 3D yarn structure of a physically aligned crocheted sample. 
The closest in output, Storck et al.'s topology model~\cite{crochet_topology_22}, also produces a continuous yarn curve, but synthesizes it from a fixed per-stitch key-point template with tension as a parametric scaling, remaining as a synthetic model rather than a reconstruction of a real crocheted object.

\subsection{Image-Based Yarn and Strand Reconstruction}
\label{subsec:image_recon}
Image-based methods recover yarn or strand structure directly from photographs or microscopy of real samples.
For woven fabrics, an extensive line of inverse-rendering work 
recovers procedural yarn parameters together with appearance properties from photographs~\cite{woven_capture_image_15, 
woven_match_micro_15, woven_capture_image_17, woven_capture_photo_22, 
woven_capture_pair_24, woven_capture_auto_25}, and a recent extension 
reaches fiber-level detail from a single microscopic image~\cite{woven_capture_fiber_26}.
For knit fabrics, image-to-instruction pipelines infer the discrete stitch label at each cell of an assumed grid~\cite{knit_inverse_neural_19, knit_inverse_proc_19}. 
For hair, strand-accurate multi-view capture reconstructs many independent 3D strands from a calibrated multi-view-stereo setup~\cite{hair_capture_strand_19}. 
Fabric pipelines require a regular grid that fixes where each yarn or stitch can lie, whereas strand pipelines assume independent, externally visible filaments emerging from a known support surface.
Neither holds for crochet, whose interlocking loops occlude most of the interior yarn from any single viewpoint (Fig.~\ref{fig:ct_input}(a)).

\begin{figure}[t]
  \centering
  \includegraphics[width=\linewidth]{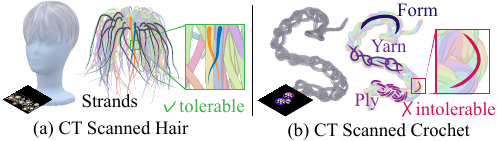}
  \caption{\textbf{Hair vs.\ crochet: hierarchy and error tolerance.}
  (a) Hair has independent strands and
  tolerates locally-plausible errors. (b) Crochet has three 
  levels (form, yarn, plies). Because the entire object is 
  constructed from a single continuous yarn, a local 
  reconstruction error will propagate globally.}
  \label{fig:yarn_vs_hair}
  \vspace*{-0.4cm}
\end{figure}

\subsection{CT-Based Yarn and Strand Reconstruction}
\rev{In textile engineering, Gong et al.~\cite{cloth_ct_xsection_09} modeled yarn cross-sections in plain woven fabric from micro-CT scans.}
Shinohara et al.~\cite{cloth_ct_extract_10} pioneered CT yarn centerline extraction on regular knit and woven fabrics. Later work probes single-yarn microstructure~\cite{cloth_ct_synchro_24}, manufacturing differences~\cite{cloth_ct_yarn_26}, and woven appearance modeling in graphics~\cite{woven_ct_appearance_11,
woven_match_micro_15}. 
These works assume regular grids or isolated straight yarns, enabling per-yarn tracing without contact disambiguation.
Closer to our setting, Zhao et al.~\cite{woven_ct_proc_16} fit a procedural yarn model to CT scans of yarn samples, recovering statistical parameters of fiber arrangement within a ply. 
Their input is short parallel yarn segments, and their output describes a yarn material rather than the path of any specific yarn through a fabric.
On the other hand, CT2Hair~\cite{hair_ct_recon_23} reconstructs independent hair strands from a CT volume. 
Hair strands are rooted on a scalp and remain largely separated, so reconstruction reduces to tracing individual strands from their roots.
A crocheted object instead consists of a single continuous yarn of twisted plies whose loops repeatedly come into close contact, so local fiber orientations do not reveal the yarn trajectory and neighboring passes are hard to distinguish in the volume.
The recovered centerline must moreover form a single continuous yarn path consistent with the physical crochet structure, a domain gap that prevents hair-reconstruction methods from being applied directly.

\begin{figure*}[t]
  \centering
  \begin{overpic}[width=\textwidth]{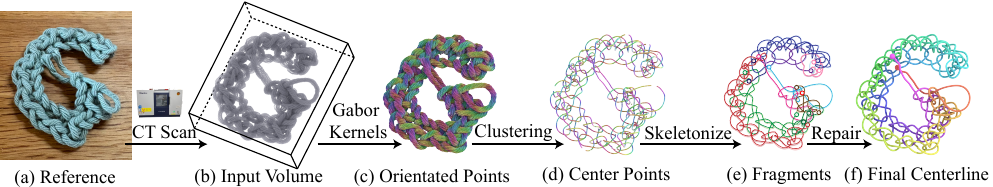}
    \put(19,15){\LARGE $\mathcal{V}$}
    \put(38,15){\LARGE $\mathcal{P}$}
    \put(55,15){\LARGE $\hat{\mathcal{P}}$}
    \put(72.5,15){\LARGE $\mathcal{F}$}
    \put(87,15){\LARGE $\mathcal{C}$}
    \put(16.5,3){\makebox(0,0){\small (Sec.~\ref{subsec:preprocess})}}
    \put(36,3){\makebox(0,0){\small (Sec.~\ref{subsec:orientation})}}
    \put(52,3){\makebox(0,0){\small (Sec.~\ref{subsec:meanshift})}}
    \put(69.5,3){\makebox(0,0){\small (Sec.~\ref{subsec:topology})}}
    \put(84.5,3){\makebox(0,0){\small (Sec.~\ref{subsec:manual})}}
  \end{overpic}
  \caption{\textbf{Pipeline overview.} (a) Reference image of the
  physical crocheted sample. (b) Micro-CT volume $\mathcal{V}$.
  (c) Per-voxel orientation field from 3D Gabor responses, shown as the
  oriented point cloud $\mathcal{P}$. (d) Center points $\hat{\mathcal{P}}$
  after anisotropic mean-shift. (e) Yarn fragments $\mathcal{F}$ extracted from the
  center points. (f) Final single continuous centerline $\mathcal{C}$ after
  \rev{automatic repairing and interactive completion}.}
  \label{fig:pipeline}
\end{figure*}

\section{Method}
\label{sec:methods}

Given a density volume $\mathcal{V}\in\mathbb{R}^{H\times W \times L}$ from a micro-CT scan, our goal is to recover a single continuous yarn path $\mathcal{C}=(\mathbf{p}_1,\ldots,\mathbf{p}_N)$, an ordered sequence of center points $\mathbf{p}_i\in\mathbb{R}^3$ whose ordering encodes the topology of the entire yarn. 
Our pipeline consists of three stages (Fig.~\ref{fig:pipeline}).
\emph{Orientation estimation} (Sec.~\ref{subsec:orientation}) assigns a fiber direction vector to each voxel above a density threshold, resulting in a sparse oriented point cloud
$\mathcal{P}$. 
\emph{Center-point extraction} (Sec.~\ref{subsec:meanshift}) applies anisotropic  
mean-shift clustering to $\mathcal{P}$ to recover yarn centerline candidates $\hat{\mathcal{P}}$.
\rev{\emph{Topology skeletonization} (Sec.~\ref{subsec:topology}) groups $\hat{\mathcal{P}}$ into ordered yarn fragments $\mathcal{F}$ and assembles them into the final single continuous yarn path $\mathcal{C}$.}
We additionally provide an interactive sketch interface for manually resolving ambiguous fragment connections when the automatic reconstruction is insufficient \rev{(Sec.~\ref{subsec:manual})}.
The recovered yarn path $\mathcal{C}$ is then used in downstream applications including simulation, rendering, and stitch-pattern extraction.

\subsection{Preprocessing}
\label{subsec:preprocess}

A micro-CT scanner reconstructs a density volume $\mathcal{V}$ from X-ray projections taken around the sample~\cite{img_recon_fdk_84}. 
Because yarn is a soft fibrous material with weak X-ray attenuation, its boundary with air is not sharply defined.
Consequently, the reconstructed density field exhibits substantial blur and noise, together with a low-intensity halo surrounding the yarn, as also observed by CT2Hair~\cite{hair_ct_recon_23}.
To suppress this haze, we threshold the non-zero voxels with Otsu's criterion~\cite{img_threshold_otsu_79}, which picks one threshold $\tau$ per volume automatically.
We avoid additional smoothing in order to preserve fine boundary structures, and use the thresholded volume in all subsequent stages.

\subsection{Orientation Field Estimation}
\label{subsec:orientation}

\begin{figure}[b]
 \centering
 \begin{overpic}[width=\linewidth]{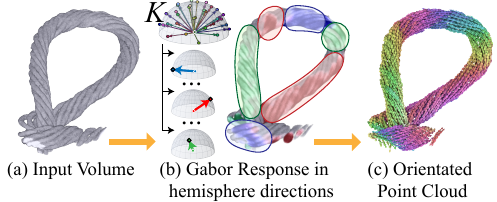}
  \put(4,36){\Large $\mathcal{V}$}
  \put(75,36){\Large $\mathcal{P}$}
  \put(28,28){ \textcolor[HTML]{0071BC}{$\mathbf{d}_1$}}
  \put(28,21){ \textcolor[HTML]{C1272D}{$\mathbf{d}_2$}}
  \put(28,14){ \textcolor[HTML]{009245}{$\mathbf{d}_3$}}
 \end{overpic}
 \caption{\textbf{Orientation field estimation.} (a) Input CT volume.
 (b) Regions running in different directions respond to their respective
 Gabor kernels, shown here for three directions. (c) Oriented point cloud
 $\mathcal{P}$, colored by tangent direction.}
 \label{fig:orientation}
\end{figure}

This stage assigns a tangent direction to every voxel inside the thresholded volume and extracts a sparse, oriented point cloud over
regions where the local structure is reliably line-like. 
Hair-CT pipelines such as CT2Hair~\cite{hair_ct_recon_23} estimate orientations from voxel intensity gradients. 
This fits hair's structure as presented in Fig.~\ref{fig:yarn_vs_hair}a: individual strands sit in air with clear boundaries, so each voxel's gradient encodes the local strand direction.
Crochet has no such clean boundaries, so gradients are unreliable. 
We instead estimate orientation with a Gabor structure-tensor formulation~\cite{img_structure_knutsson_11}: each 3D Gabor kernel is
bandpass at the yarn scale, so it responds to the whole yarn-scale ridge rather than to a boundary. 
We sample $K$ carrier directions $\mathbf{d}_k$ uniformly on the upper hemisphere via a Fibonacci lattice (Fig.~\ref{fig:orientation}b). 
For each direction, we construct a 3D Gabor kernel whose width is determined by the yarn radius $R$ (measured from a CT slice) and whose carrier wave is aligned with $\mathbf{d}_k$.
We evaluate the convolutions efficiently in the frequency domain using FFTs.
The squared response $|r_k(\mathbf{x})|^2$ is large  when the local fiber orientation aligns with $\mathbf{d}_k$ at $\mathbf{x}$. 
Summing over all directions gives the per-voxel structure tensor
\begin{equation}
\mathbf{T}(\mathbf{x}) = \sum_{k=1}^{K} |r_k(\mathbf{x})|^2 \,
\mathbf{d}_k \mathbf{d}_k^{\top},
\end{equation}
The eigendecomposition of $\mathbf{T}$ then reveals the dominant local fiber orientation.
In a locally linear neighborhood, one eigenvalue dominates and the smallest eigenvector aligns with the principal fiber direction.
We take this smallest eigenvector as the per-voxel local tangent direction $\mathbf{t}(\mathbf{x})\in\mathbb{R}^3$, where $|\mathbf{t}(\mathbf{x})|=1$.

\rev{We then keep only voxels with a strong enough Gabor response. The mean per-direction response $\bar{r}(\mathbf{x}) = \tfrac{1}{K}\sum_k |r_k(\mathbf{x})|$ is large inside fiber material and small in air, background, or the surrounding haze, so we drop voxels whose mean response falls below $\tau_E$. Averaging over the $K$ directions makes $\tau_E$ independent of the number of Gabor directions.}
Finally, we reduce the density of the remaining samples by voxel binning, keeping one point per cubic cell of side length $d$ times the mean voxel spacing.
The resulting sparse oriented point cloud is denoted by $\mathcal{P}=\{(\mathbf{p}_i,\mathbf{t}_i)\}$.

\subsection{Center Point Set Extraction}
\label{subsec:meanshift}

\begin{figure}[t]
 \centering
 \begin{overpic}[width=\linewidth,percent]{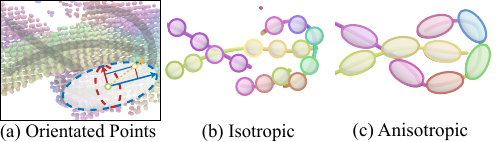}
  \put(15,10){\Large \textcolor[HTML]{009245}{$\mathbf{p}_i$}}
  \put(27,18){\Large \textcolor[HTML]{ffbb56}{$\mathbf{p}_j$}}
  \put(24.5,5.5){\Large\textcolor[HTML]{0071BC}{$\delta_{tan}$}}
  \put(16,18){\Large \textcolor[HTML]{C1272D}{$\delta_{perp}$}}
 \end{overpic}
 \caption{\textbf{Anisotropic mean-shift kernel.} (a) Oriented point cloud
 $\mathcal{P}$. (b) An isotropic kernel fuses neighboring passes where they
 run close together. (c) Our prolate kernel, elongated along the tangent,
 reaches far along the yarn axis and preserves its structure.}
 \label{fig:aniso_kernel}
\end{figure}

The points in $\mathcal{P}$ are spread across the yarn volume, not on its centerline. 
We therefore iteratively project the point set toward the yarn centerline using a mean-shift procedure, which repeatedly moves each point toward the weighted average of its neighbors.
Unlike conventional isotropic mean-shift, our method is anisotropic and exploits the local tangent direction at each point.
First, neighboring points are weighted using a kernel elongated along the local tangent direction $\mathbf{t}_i$ (Fig.~\ref{fig:aniso_kernel}), so that the neighborhood extends farther along the yarn than across it.
Second, the mean-shift update is projected onto the plane perpendicular to $\mathbf{t}_i$, preventing points from drifting along the yarn while allowing them to move toward the centerline.

Specifically, we define an anisotropic Gaussian kernel weight between points $\mathbf{p}_i$ and $\mathbf{p}_j$.
Let $\mathbf{p}_j-\mathbf{p}_i$ be their relative offset.
Decomposing this offset with respect to the local tangent
$\mathbf{t}_i$ gives its component along the tangent,
$\delta_{\mathrm{tan}} = (\mathbf{p}_j -
\mathbf{p}_i) \cdot \mathbf{t}_i$, and its component perpendicular to the tangent,
$\delta_{\mathrm{perp}} = \|(\mathbf{p}_j - \mathbf{p}_i) - \delta_{\mathrm{tan}}
\mathbf{t}_i\|$. 
The kernel weight is defined as
\begin{equation}
w_{ij} = \exp\!\left(
 -\frac{\delta_{\mathrm{tan}}^{2}}{2\sigma_{\mathrm{tan}}^{2}}
 - \frac{\delta_{\mathrm{perp}}^{2}}{2\sigma_{\mathrm{perp}}^{2}}
\right)
\,\bigl|\mathbf{t}_i \cdot \mathbf{t}_j\bigr|^{\alpha}.
\label{eq:aniso_kernel}
\end{equation}
The anisotropic bandwidths $\sigma_{\mathrm{tan}} = \rho h$ and $\sigma_{\mathrm{perp}} = h$  make the kernel elongated along the yarn direction, extending 
$\rho>1$ times farther along the tangent than across it.
The trailing factor up-weights neighbors whose tangent agrees with $\mathbf{t}_i$. 
The parameter settings are listed in Table~ \ref{tab:notation_params}.

We then move each point toward the weighted mean of its neighbors, with the shift projected onto the plane perpendicular to $\mathbf{t}_i$ so that only
off-axis motion is applied: 
\begin{equation}
\mathbf{p}_i \;\leftarrow\; \mathbf{p}_i
+ \bigl(\mathbf{I} - \mathbf{t}_i \mathbf{t}_i^{\top}\bigr)
\!\left(
 \frac{\sum_j w_{ij}\, \mathbf{p}_j}{\sum_j w_{ij}} - \mathbf{p}_i
\right).
\label{eq:ms_update}
\end{equation}
Here the parenthesized term is the mean-shift vector toward the
kernel-weighted average of the neighbors, and the projection
$\mathbf{I} - \mathbf{t}_i\mathbf{t}_i^{\top}$ keeps only its component perpendicular to $\mathbf{t}_i$, so each point moves toward the centerline without drifting along the yarn.
After each shift, we update the tangent direction by averaging neighboring tangents and renormalizing the result,
\begin{equation}
\mathbf{t}_i' = \frac{\sum_j w_{ij}\,\mathbf{t}_j}{\sum_j w_{ij}},\;\;\;
\mathbf{t}_i \leftarrow \frac{\mathbf{t}_i'}{\|\mathbf{t}_i'\|}
\label{eq:tangent_refresh}
\end{equation}
which keeps the kernel aligned with the local yarn direction as the cloud tightens. 
Because the tangent represents an unoriented local direction field, its sign is arbitrary.
Before evaluating the weighted average in \eqref{eq:tangent_refresh}, we therefore flip the sign of $\mathbf{t}_j$ whenever
$\mathbf{t}_j\cdot\mathbf{t}_i<0$.
Near endpoints and gaps, where the neighborhood becomes strongly one-sided, we attenuate the tangent update and retain the previous tangent estimate.

A conventional mean-shift procedure attracts points toward local density maxima.
However, when applied to samples distributed along a yarn, this density-driven aggregation tends to collapse neighboring points into a small number of dense clusters, destroying the spatial ordering needed to recover the centerline.
The neighborhood relations are still reliable during the early iterations, before substantial aggregation occurs.
We therefore freeze these relations at an early stage and preserve them throughout the optimization 
using springs that penalize both compression and stretching relative to the distances recorded at the freeze stage.
This regularization keeps neighboring points close to one another and prevents the centerline candidates from collapsing into disconnected clusters.

We progressively anneal the bandwidth, decreasing $h$ geometrically from $h_0$ to $h_T$ over $T$ iterations. Because $\sigma_{\mathrm{tan}}$ and $\sigma_{\mathrm{perp}}$ are both proportional to $h$, they anneal jointly while their ratio $\rho$ stays fixed: a large initial bandwidth pulls scattered points
across noise and intermediate gaps, and the smaller final bandwidth tightens the cloud onto a thin centerline. 
The inner per-point loop of this procedure is data-parallel because each $\mathbf{p}_i$ reads only from its precomputed neighbors and writes only its own slot, so we run it on the GPU and complete a full mean-shift iteration over the entire cloud in milliseconds.

\subsection{\rev{Automatic Topology Skeletonization}}
\label{subsec:topology}

\paragraph*{Fragment Extraction and Junction Detection}
The center points $\hat{\mathcal{P}}$ from mean-shift form an unordered cloud.
We therefore organize them into ordered polylines. 
We first construct a proximity graph on $\hat{\mathcal{P}}$ by connecting pairs of points whose distance is below radius $r_C$, and discard connected components smaller than $n_{\min}$ points as residual noise. 
Within each remaining component we compute a minimum spanning tree (MST) and extract its diameter, i.e., the longest path between two leaves. 
This path typically follows the dominant yarn trajectory through the component and provides an ordering of its points.
A component does not always correspond to a single clean yarn segment. 
MST nodes of degree $\ge 3$ appear precisely where two yarn passes run close enough that the proximity graph links them across the contact, and the local structures they create are the ring, twig, and collapse patterns resolved in the junction cleaning and loop detection below. 
We therefore flag every such node as a junction. 
The output is a set of ordered fragments $\mathcal{F} = \{\mathcal{F}_1, \ldots, \mathcal{F}_m\}$, each of which is an ordered polyline, together with the flagged junctions.

\paragraph*{Fragment Linking}
The fragments themselves are still disconnected across the yarn-to-yarn gaps.
To bridge them, we grow an ellipsoid at each fragment endpoint, elongated along the endpoint tangent so that it reaches farther along the yarn than across it\rev{, with its cross-section set by the proximity radius $r_C$}.
\rev{When a pair of endpoint ellipsoids overlap and neither end has any other overlapping partner, the connection is the unique local candidate and we link the two fragments directly.}
\rev{Otherwise we keep each overlapping pairing only as a candidate bridge.}

\begin{figure}[t]
 \centering
 \begin{overpic}[width=\linewidth,percent]{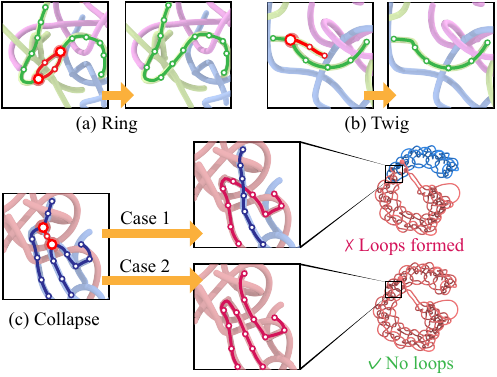}
  \put(11.5,34){\textcolor[HTML]{434597}{$a$}}
  \put(4,20){\textcolor[HTML]{434597}{$b$}}
  \put(13.5,19){\textcolor[HTML]{434597}{$c$}}
  \put(13.5,27){\textcolor[HTML]{434597}{$d$}}

  \put(50,44){\textcolor[HTML]{434597}{$a$}}
  \put(43,30){\textcolor[HTML]{c12b5b}{$b$}}
  \put(52.5,29){\textcolor[HTML]{434597}{$c$}}
  \put(52,37){\textcolor[HTML]{c12b5b}{$d$}}

  \put(50,19){\textcolor[HTML]{c12b5b}{$a$}}
  \put(43,5){\textcolor[HTML]{c12b5b}{$b$}}
  \put(52.5,4){\textcolor[HTML]{c12b5b}{$c$}}
  \put(52,12){\textcolor[HTML]{c12b5b}{$d$}}

  \put(22.5,26){\tiny\textcolor[HTML]{000000}{$\{(a,c), (b,d)\}$}}
  \put(22.5,16.6){\tiny\textcolor[HTML]{000000}{$\{(a,d), (b,c)\}$}}
 \end{overpic}
 \caption{\textbf{Three local patterns automatically resolved by our topology reconstruction.} (a) \textit{Ring}: a small closed loop inside a fragment.
 (b) \textit{Twig}: a short spurious branch. (c) \textit{Collapse}: two
 near-parallel passes wrongly merged.}
 \label{fig:junction_cases}
\end{figure}

\paragraph*{Junction Cleaning}
The flagged junctions from the fragment extraction fall into three local geometric patterns (Fig.~\ref{fig:junction_cases}).
A \textit{ring} is a small closed loop inside a fragment, where two junction nodes are joined by two distinct paths.
We resample both paths and average their corresponding points into a single centerline that replaces the pair (Fig.~\ref{fig:junction_cases}a).
A \textit{twig} is a short spurious branch at a degree-three node: of the three
branches meeting there, the one far shorter than the main path is pruned
(Fig.~\ref{fig:junction_cases}b).
Both patterns are resolved locally and without ambiguity.
A \textit{collapse} is where two near-parallel passes meet, forming a junction with four incident arms. We label them $a$, $b$, $c$, $d$, with $a,b$ on one side of the junction and $c,d$ on the other. A through-pass can never join two arms of the same side, so the four arms can be reconnected in only two ways, either $\{(a,c), (b,d)\}$ or $\{(a,d), (b,c)\}$, shown as Case~1 and Case~2 in Fig.~\ref{fig:junction_cases}c.
\rev{The same-side pairing $\{(a,b), (c,d)\}$ is excluded because a collapse arises only where two near-parallel passes pinch together. A yarn that actually turned back to join $a$ and $b$ would form a single smooth curve on that side and would not produce a four-arm junction of this shape.}
Unlike the ring and twig, this choice cannot be made from local geometry alone, so we keep the collapse as an undecided connection with two cases.

\paragraph*{Loop Detection}
To resolve the remaining undecided connections, the candidate bridges and the collapse cases, we exploit the single-strand structure of crochet.
A physically crocheted sample is wound from one continuous strand, so its true centerline is topologically an open path: one connected component, two endpoints, and no cycle. 
Any cycle in the reconstructed topology therefore implies a wrong connection, since the correct
centerline cannot revisit a point along its own trajectory. 
Rather than resolving ambiguous junctions locally, we retain all possible connection choices.
Each collapse contributes two cases and each candidate bridge two states, so $N$ undecided connections give $2^{N}$ candidate topologies. 
We enumerate them and select the unique configuration that forms a single open path as the ordered yarn path $\mathcal{C}$.
Only the correct configuration keeps every fragment on one continuous centerline without a loop, which for a collapse means all four arms $a$, $b$, $c$, $d$ stay on the same strand
(Fig.~\ref{fig:junction_cases}c).
\rev{The enumeration is tractable in practice: $N$ peaks at five across our dataset, and branches are pruned early because a single closed cycle already invalidates every configuration that contains it.}
\rev{When no candidate forms a single open path, a wrong connection must have been introduced by an earlier step, which we cannot resolve automatically, so all offending junctions are passed to the manual repair stage (Sec.~\ref{subsec:manual}).}
\rev{Once the topology is fixed, automatically or after manual repair, we fit a B-spline of smoothing factor $s_C$ through its ordered center points to obtain the smooth final centerline.}

\begin{figure}[b]
 \centering
 \includegraphics[width=\linewidth]{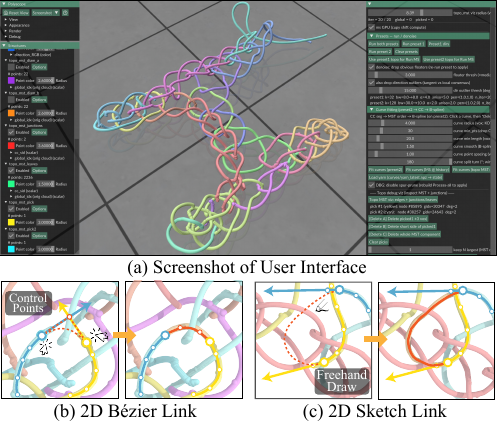}
 \caption{\textbf{Manual editing interface.} (a) Topology graph overlaid on
 the 3D view. (b) B\'ezier mode: the user shapes a smooth path between two
 picked endpoints with control handles. (c) Sketch mode: the user draws a
 freehand stroke between them. Either 2D curve is back-projected to 3D as a
 bridge.}
 \label{fig:sketch}
\end{figure}

\subsection{\rev{Manual Repair}}
\label{subsec:manual}

\rev{While the automatic stage is designed to recover the complete yarn path, crochet regions with complex topology can still produce highly ambiguous breaks or missing connections that no automatic rule resolves reliably.}
\rev{As a fallback for these cases, we provide a lightweight interactive interface built on \texttt{Polyscope}~\cite{viz_polyscope_19} for manual repair, offering two ways to connect two broken ends, a B\'ezier link and a freehand sketch.}
As Fig.~\ref{fig:sketch}a shows, the user sees a 3D view of the current topology.
The user picks two nodes in the current 3D view, and we fit a 2D B\'ezier curve between them in screen space. 
Given the tangents at the two picked nodes, the control points can be computed by intersecting these tangents, as Fig.~\ref{fig:sketch}b shows, and the user can also adjust them to specify a smooth 2D path between the two endpoints.
Alternatively, for connecting nodes that are far apart, the user can directly sketch a
freehand stroke between the two picks in screen space, as Fig.~\ref{fig:sketch}c shows.
Each pixel along that 2D curve is back-projected to 3D by linearly interpolating its depth between the two picks, yielding a bridge curve that passes through both endpoints, matches the drawn shape from the chosen viewpoint, and is inserted into the topology graph as an additional segment.
After user finish all edits, we rerun the loop detection of the automatic stage on the updated topology, verifying that the repaired centerline remains a valid single open path.

\section{Experiments}
\label{sec:exp}

\begin{table}[t]
  \centering
  \footnotesize
  \setlength{\tabcolsep}{4pt}
  \begin{tabular}{@{}llr@{}}
    \toprule
    Symbol & Meaning & Value \\
    \midrule
    \multicolumn{3}{@{}l@{}}{\textit{Input}} \\
    $\mathcal{V}$ & CT volume & --- \\
    \midrule
    \multicolumn{3}{@{}l@{}}{\textit{Orientation field}} \\
    $K$ & Gabor carrier directions (Fibonacci) & $128$ \\
    $\sigma_{\mathrm{Gabor}}$ & Gabor envelope and tangent-smoothing width (vox) & $7.0$ \\
    $\omega$ & Gabor frequency & $0.05$ \\
    $\tau_E$ & \rev{Per-direction response threshold} & \rev{$250$} \\
    $\mathbf{t}(\mathbf{x})$ & Per-voxel tangent vector & --- \\
    $\mathcal{P}$ & Oriented point cloud & --- \\
    \midrule
    \multicolumn{3}{@{}l@{}}{\textit{Center point set extraction}} \\
    $\rho$ & Anisotropy ratio & $2.0$ \\
    $\alpha$ & Direction power & $2$ \\
    $h_0, h_T$ & Bandwidth schedule (vox) & $30 \rightarrow 10$ \\
    $T$ & Mean-shift iterations & $20$ \\
    $\sigma_{\mathrm{tan}}$ & Kernel width along tangent & $\rho h$ \\
    $\sigma_\mathrm{perp}$ & Kernel width across tangent & $h$ \\
    $\hat{\mathcal{P}}$ & Center points (post mean-shift) & --- \\
    \midrule
    \multicolumn{3}{@{}l@{}}{\textit{Topology Skeletonization \& Manual Repair}} \\
    $r_C$ & Proximity radius (vox) & $4.0$ \\
    $n_{\min}$ & Min.\ component size (pts) & $30$ \\
    $s_C$ & B-spline smoothing factor & $1.5$ \\
    $\mathcal{F}$ & Yarn fragments  & --- \\
    $\mathcal{C}$ & Output yarn curves  & --- \\
    \midrule
    \multicolumn{3}{@{}l@{}}{\textit{Stitch fitting}} \\
    $M$ & Sample points per stitch & $30$ \\
    $\mathcal{S}_t$ & Canonical stitch template & --- \\
    $\mathcal{S}_c$ & Candidate stitch (centerline window) & --- \\
    $e^{*}$ & Refined stitch end point & --- \\
    \bottomrule
  \end{tabular}
  \caption{Key symbols used in the method together with the
  parameter values used in our experiments.  The same configuration
  is applied to every sample across the whole dataset.}
  \label{tab:notation_params}
\end{table}
\begin{figure}[b]
 \centering
 \includegraphics[width=\linewidth]{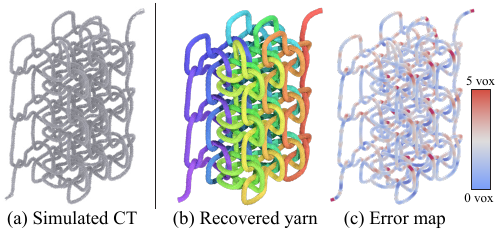}
 \caption{\rev{\textbf{Validation on synthetic data.} (b) The centerline our
 pipeline recovers from the synthetic volume (a), colored by arc length, and
 (c) its distance to the exact ground truth, colored from $0$ to
 $5$ voxels.}}
 \label{fig:synthetic}
\end{figure}

\paragraph*{Implementation Details}
The pipeline is implemented in Python with GPU kernels in \texttt{PyTorch}.
All measurements reported in this section are performed on a workstation with an AMD~Ryzen~9~9950X (16-core /
32-thread, 5.0\,GHz), 128\,GB of system memory, and a single NVIDIA~GeForce~RTX~3090 (24\,GB VRAM).
Table~\ref{tab:notation_params} lists the symbols used throughout the method together with the numerical values we use in our experiments.
The same set is used for every sample unless otherwise noted.
\rev{Most parameters are physically grounded. The yarn radius $R$ read from a CT slice sets the Gabor width $\sigma_{\mathrm{Gabor}}$, frequency $\omega$, and mean-shift bandwidths $h_0\!\rightarrow\!h_T$, so adapting to a different sample or scanner mainly amounts to re-measuring the yarn radius voxels. The response threshold $\tau_E$ is defined per Gabor direction and is thus independent of the number of directions $K$. The remaining parameters, those of the center point extraction and topology stages, are chosen empirically. Regarding the choice of $K$, the recovered geometry is insensitive to it, with the Chamfer distance varying within $0.13$ voxels for $K$ from $32$ to $256$. Larger $K$ sharpens the tangents at higher cost, and we use $K{=}128$ as the trade-off.}

\subsection{Dataset}
\label{sec:dataset}

\begin{figure*}[t]
  \centering
  \makebox[\textwidth][r]{%
  \begin{overpic}[width=\textwidth]{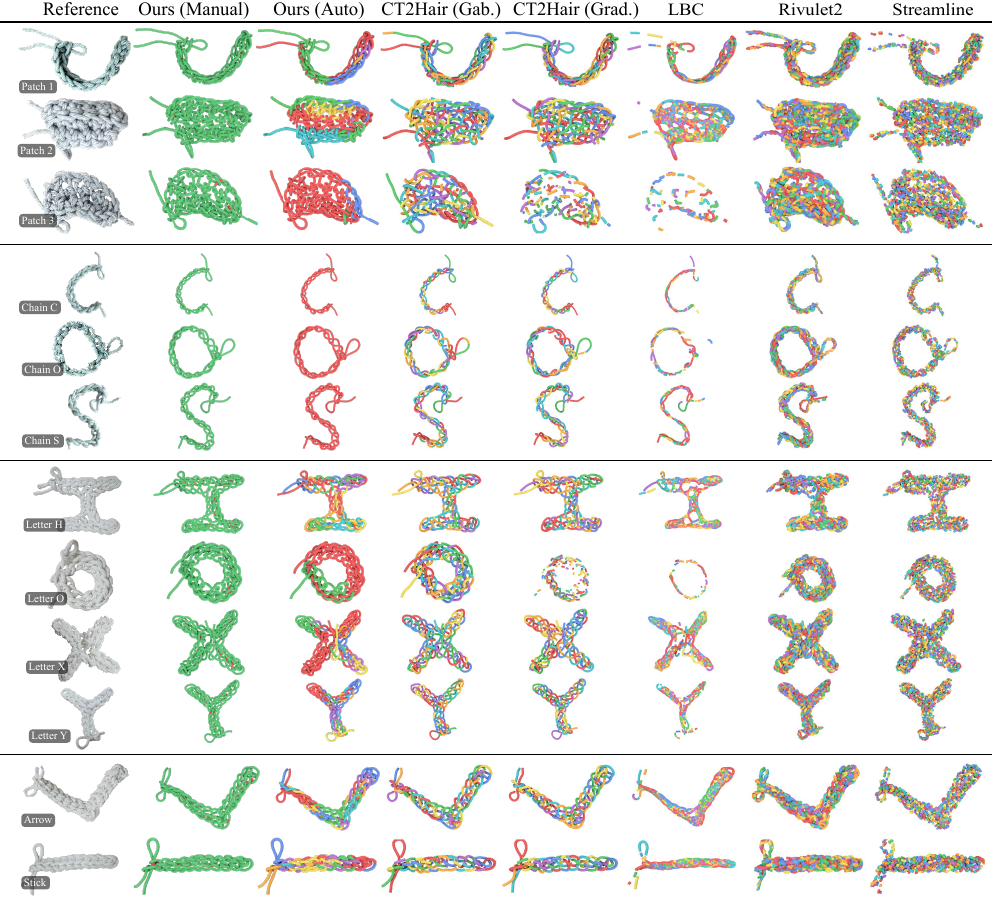}
    \put(-0.3,77.6){\rotatebox{90}{\makebox[0pt]{\textsc{Patches}}}}
    \put(-0.3,54.9){\rotatebox{90}{\makebox[0pt]{\textsc{Chain~stitches}}}}
    \put(-0.3,29.8){\rotatebox{90}{\makebox[0pt]{\textsc{Letters}}}}
    \put(-0.3,7.6){\rotatebox{90}{\makebox[0pt]{\textsc{Misc}}}}
  \end{overpic}}
\caption{\textbf{Visual comparison against five baselines on twelve samples grouped by category.}
\rev{Reference shows a photograph of the physical sample. In Ours (Manual), green marks the automatically recovered yarn and red the parts completed by interactive sketch edits. All other columns are colored by fragment identity, one color per recovered curve.}
}
\label{fig:comparison}
\end{figure*}

\paragraph*{Acquisition.}  We collected a dataset of 18 crocheted samples imaged on a Rigaku HX~Lab130 high-resolution micro-CT scanner.
Each volume is reconstructed at full resolution, spanning roughly $900$ to $2200$ voxels in the two in-plane axes and $300$ to $550$ slices along the scan axis, and stored as a single 16-bit Nearly Raw Raster Data (NRRD) file.
To isolate geometric variability from material variability, all samples are crocheted from the same recycled cotton yarn intended for $3.0$--$4.0$~mm crochet hooks.
\rev{We additionally synthesize a volume with a known centerline by simulating a yarn model along the single-yarn path of the $3{\times}3{\times}3$ solid-knit cube of Hirose et al.~\cite{hirose:2024:solid-knitting}. The model twists eight plies of $40$ fibers each into a yarn of radius $15$ voxels, which we rasterize into a density volume and degrade with a Gaussian point-spread function, distinct air and yarn intensity distributions, and spatially correlated noise, so that the contrast between touching yarn passes is as weak as in our real scans.}

\paragraph*{Composition.}  The 18 samples fall into five categories
spanning the structural variety in real crochet pieces:
\textsc{Patches} (3, single-stitches crocheted patches), 
\textsc{Chain~stitches} (3, the
simplest chain stitch patterns), 
\textsc{Letters} (6, stitched letter forms with
more complex junctions), \textsc{Numbers} (4, including the digit
two at three different tensions and the digit six)\footnote{The
crocheted \textsc{Letters} and \textsc{Numbers} patterns follow designs by
amimozi, \url{https://amimozi.myportfolio.com/}.},
and \textsc{Misc} (2, stick and arrow shaped crochet). The number in brackets refers to the number of samples in each category.
\subsection{Metrics}
\label{sec:metrics}

\rev{The goal of reconstruction is a yarn path that is visually and topologically consistent with the scanned object, since the downstream applications depend on this overall consistency rather than on sub-voxel geometric agreement. Our evaluation therefore rests primarily on visual comparison (Fig.~\ref{fig:comparison}), with quantitative metrics as supporting evidence.}
\rev{As no ground-truth yarn geometry is available for real crocheted samples, the metrics are computed against a \emph{reference} yarn path that we manually annotate for each sample using the interactive editor, and all results are averaged over the 18 samples.}
\rev{Two metrics are \emph{reference-free}: the number of recovered curves $\#\mathcal{C}$ (ideally $1$, a single continuous yarn) and the mean recovered-curve arc-length $\overline{L_{\mathcal{C}}}$ in voxels, which together describe how fragmented the output is. The remaining four are \emph{reference-based}: the relative total-length error $E_L = |L/L_{\mathrm{ref}} - 1|$, the bidirectional Chamfer distance $\ell_{\mathrm{CD}}$ between the recovered and reference point sets, the mean distance $\ell_2$ from each reference point to its nearest recovered point, both in voxels, and the median angle $\widetilde{\alpha}$ between matched tangents. Each metric is blind to a different failure mode, so no single number should be read in isolation.}

\begin{figure*}[t]
 \centering
 \includegraphics[width=\textwidth]{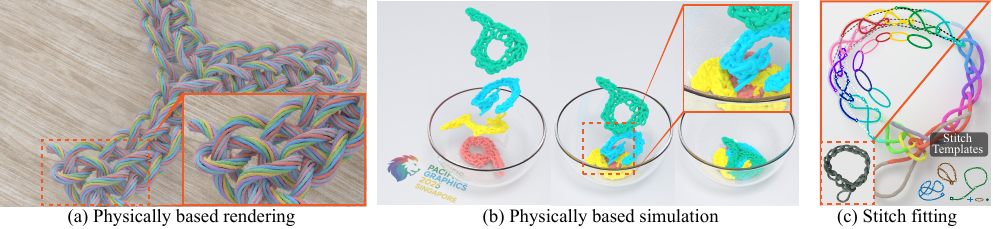}
 \caption{\textbf{Downstream applications of the recovered yarn
 path.} (a) Physically based ply-level rendering in \texttt{Mitsuba}~3.
 (b) Physically based simulation as a Discrete Elastic Rod.
 (c) Stitch pattern fitting via template matching.}
 \label{fig:applications}
\end{figure*}

\subsection{Validation on Synthetic Data}
\label{sec:synthetic}

\rev{To complement the manually annotated references with an absolute ground truth, we run our pipeline on the synthetic volume described in Sec.~\ref{sec:dataset}, whose exact centerline is known by construction. The automatic stage alone recovers a single continuous centerline here, without any interactive edit, at a total-length error of $1.1\%$ and a Chamfer distance of $3.45$ voxels, or $0.23$ yarn radii. As Fig.~\ref{fig:synthetic} shows, the deviation stays below a third of the yarn radius along $95\%$ of the path and grows only at the two free yarn ends, where the reconstruction stops short of the true tips. The accuracy observed against our manual annotations is therefore consistent with the accuracy against an exact ground truth.}

\subsection{Comparison to Baselines}

We compare against five baselines on our crochet dataset.  
Two are CT2Hair-derived pipelines~\cite{hair_ct_recon_23}: \emph{Gabor} feeds our Gabor-oriented point cloud into the mean-shift stage of CT2Hair, while \emph{Gradient} keeps that pipeline but replaces the Gabor field with the smoothed gradient field of CT2Hair.  
The remaining three operate on different principles.  
Laplacian-based contraction~\cite{5521461} contracts the point
cloud onto a one-dimensional skeleton by repeated Laplacian smoothing.  
The Rivulet2 neuron tracer~\cite{liu2018automated} back-traces tubular structures directly from the volume along intensity ridges.
Streamline tracing integrates curves by following the dense per-voxel orientation field.

\rev{Figure~\ref{fig:comparison} shows the comparison on twelve samples. Our reconstruction reproduces the yarn structure faithfully on every sample, whereas each baseline exhibits a characteristic failure such as fragmentation, missing arcs, parallel-rail duplication, or over-contraction. The quantitative results in Table~\ref{tab:metrics_main}, averaged over all 18 samples, agree with this visual impression. Our automatic pipeline returns by far the fewest curves and the longest mean arc-length, and it leads on the reference-based metrics as well. Its average of 15 curves per sample is still above the ideal of a single curve, so the automatic output is a strong starting point rather than a finished reconstruction.}
\rev{The interactive effort that closes this gap is small. All 18 samples reach a single continuous yarn path, with a median of $4$ sketch edits and $2$:$46$ of editing (mean $7.5$ edits, $3$:$49$), and the edited strand contributes $2.0\%$ of the final centerline length on average, $1.0\%$ at the median. The three chain-stitch samples need no editing at all.}
\begin{table}[t]
  \centering
  \footnotesize
  \setlength{\tabcolsep}{2.6pt}
  \begin{tabular}{@{}l|cc|cccc|cc@{}}
    \toprule
     & \multicolumn{2}{c|}{Reference-free} & \multicolumn{4}{c|}{Reference-based} & \multicolumn{2}{c}{Effort} \\
    \cmidrule(lr){2-3}\cmidrule(lr){4-7}\cmidrule(lr){8-9}
    Method & $\#\mathcal{C}$ & $\overline{L_{\mathcal{C}}}$ & $E_L\downarrow$ & $\widetilde{\alpha}\downarrow$ & $\ell_{\mathrm{CD}}\downarrow$ & $\ell_2\downarrow$ & edits & time \\
    \midrule
    CT2Hair (Gab.) & 88 & 314 & 0.115 & 10.55 & 16.99 & 18.55 & --- & --- \\
    CT2Hair (Grad.) & 115 & 236 & 0.271 & 12.75 & 20.60 & 24.67 & --- & --- \\
    LBC & 191 & 92 & 0.456 & 53.94 & 31.92 & 44.29 & --- & --- \\
    Rivulet2 & 676 & 140 & 2.141 & 35.53 & 23.31 & 17.53 & --- & --- \\
    Streamline & 730 & 49 & 0.230 & 25.05 & 21.32 & 19.41 & --- & --- \\
    \midrule
    \textit{Ours} (Auto) & 15 & 3814 & \textbf{0.002} & \textbf{2.66} & \textbf{2.43} & \textbf{2.26} & --- & --- \\
    \textit{Ours} (Manual) & \textbf{1} & \textbf{30359} & --- & --- & --- & --- & 4 & 2:46 \\
    \bottomrule
  \end{tabular}
  \caption{\rev{\textbf{Quantitative comparison to baselines.} Averaged over the
  18 samples, with metrics defined in Sec.~\ref{sec:metrics} and arrows marking
  the better direction. Effort is the median number of sketch edits and editing
  time (min:s) per sample. Our automatic output already leads every baseline,
  and a few edits complete it into a single continuous yarn on all 18
  samples.}}
  \label{tab:metrics_main}
  \vspace*{-0.3cm}
\end{table}

\subsection{Ablation Study}
\label{sec:ablation}

\begin{table}[b]
  \centering
  \footnotesize
  \setlength{\tabcolsep}{3pt}
  \begin{tabular}{@{}l|cc|cccc@{}}
    \toprule
     & \multicolumn{2}{c|}{Reference-free} & \multicolumn{4}{c}{Reference-based} \\
    \cmidrule(lr){2-3}\cmidrule(lr){4-7}
    Variant & $\#\mathcal{C}$ & $\overline{L_{\mathcal{C}}}$ & $E_L\downarrow$ & $\widetilde{\alpha}\downarrow$ & $\ell_{\mathrm{CD}}\downarrow$ & $\ell_2\downarrow$ \\
    \midrule
    w/o Anisotropic MS         & 22 & 2511 & 0.022 & 3.46 & 3.45 & 3.14 \\
    w/o Direction Update       & 19 & 2703 & 0.024 & 3.03 & 2.85 & 2.65 \\
    w/o Topology               & 20 & 2155 & 0.003 & 2.69 & 2.54 & 2.34 \\
    \midrule
    \textit{Ours} (Full)              & \textbf{15} &  \textbf{3814} & \textbf{0.002} & \textbf{2.66} & \textbf{2.43} & \textbf{2.26} \\
    \bottomrule
  \end{tabular}
  \caption{\textbf{Center-extraction ablation.} This table shows 
  the performance of different variants of our center-extraction 
  method, averaged over all samples. Arrows mark the better direction.
  }
  \label{tab:ablation}
\end{table}

We ablate the components of the center-point extraction stage
(Sec.~\ref{subsec:meanshift}) across all 18 samples in our dataset.
\emph{Full} is the
complete center-extraction pipeline and each remaining variant 
drops one mean-shift component: the
anisotropic kernel (reverting to an isotropic ball), the per-iteration
direction update, and the topology-preservation step.
The anisotropic kernel matters most.  
Replacing it with an isotropic ball
inflates the $\ell_{\mathrm{CD}}$ and $\ell_2$ errors by roughly $40\%$,
raises the
tangent error, and fragments the cloud into half-again as many curves,
because a round kernel pulls neighboring passes together instead of
following each strand.  Dropping the per-iteration direction update degrades
the geometry more mildly, and removing topology preservation leaves the
point-wise metrics almost unchanged while raising the curve count,
confirming that its effect is mainly on connectivity rather than point
placement.

\begin{figure}[b]
 \centering
 \begin{overpic}[width=\linewidth,percent]{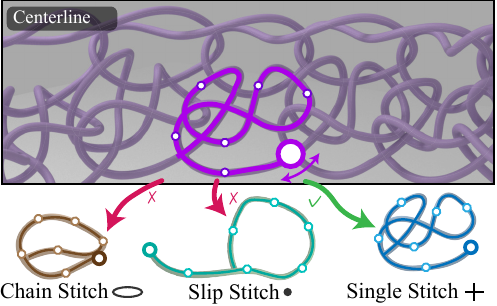}
  \put(20,56.5){\large \textcolor[HTML]{ffffff}{$\mathcal{C}$}}
  \put(46,50){\huge \textcolor[HTML]{AB00E9}{$\mathcal{S}_{\mathrm{c}}$}}
  \put(14,20){\Large \textcolor[HTML]{A67C52}{$\mathcal{S}_{\mathrm{ch}}$}}
  \put(35,20){\Large \textcolor[HTML]{4ea79d}{$\mathcal{S}_{\mathrm{sl}}$}}
  \put(71,20){\Large \textcolor[HTML]{0071BC}{$\mathcal{S}_{\mathrm{sg}}$}}
  \put(57,30){\small \textcolor[HTML]{651990}{$e^*$}}
  \put(70,12){\textcolor[HTML]{60b255}{$\mathcal{S}_{\mathrm{t}}^*$}}
 \end{overpic}
 \caption{\textbf{Stitch pattern fitting.} From a seed on the recovered
 centerline, each candidate window is rigidly aligned to three stitch-type
 templates (bottom: chain, slip, single) with the Kabsch algorithm.
 $\mathcal{S}_t^*$ is the template that yields the smallest RMSD.
 $e^*$ is the refined end point of that best-fit stitch, which becomes the
 start anchor of the next iteration.}
 \label{fig:template_fitting}
 \vspace*{-0.2cm}
\end{figure}

\subsection{Applications for the Extracted Centerline}

\paragraph*{Simulation and Rendering}
To produce ply-level geometry we wrap the recovered centerline with the coaxial-helix procedural ply model of Zhao et al.~\cite{woven_ct_proc_16}, yielding a twisted thread that follows the centerline. 
We render this geometry in \texttt{Mitsuba}~3~\cite{render_mitsuba3_22} with a custom shader that adds
Perlin noise~\cite{schroder2014image} along the ply axis for the fiber-level fuzz visible at
the yarn surface (Fig.~\ref{fig:applications}a). 
For physically based simulation we treat the centerline as a discrete elastic 
rod~\cite{sim_der_08} and integrate it under gravity and self-contact (Fig.~\ref{fig:applications}b).

\paragraph*{Stitch Pattern Extraction}
Once the centerline $\mathcal{C}$ has been recovered, we decompose it into an ordered sequence of stitches $\mathcal{S} = \{\mathcal{S}_1,\ldots,\mathcal{S}_n\}$, where each stitch $\mathcal{S}_i$ is an interval of $\mathcal{C}$ tagged with a stitch type.
The user marks where the first stitch $\mathcal{S}_1$ starts. 
We then sweep a window along the centerline and label every subsequent stitch by matching it against a small retrieval library of three canonical templates: a chain stitch, a slip stitch, and a single stitch, shown in Fig~\ref{fig:template_fitting}.
For a candidate stitch $\mathcal{S}_c$ ending at $e$, we sample $M$ points uniformly along both $\mathcal{S}_c$ and each canonical template $\mathcal{S}_t$ in the library $\{\mathcal{S}_{\mathrm{ch}},
\mathcal{S}_{\mathrm{sl}}, \mathcal{S}_{\mathrm{sg}}\}$, and rigidly align the two point sets with the Kabsch algorithm~\cite{kabsch1976} by measuring the root mean square deviation (RMSD) between them.
After alignment, the per-template residual is
\begin{equation}
\mathrm{RMSD}(\mathcal{S}_c, \mathcal{S}_t) = \sqrt{\frac{1}{M}
\sum_{k=1}^{M} \bigl\| \mathcal{S}_c(k) - \mathcal{S}_t(k) \bigr\|^{2}},
\label{eq:rmsd}
\end{equation}
where $\mathcal{S}_c(k)$ and $\mathcal{S}_t(k)$ denote the $k$-th of the $M$ aligned sample points. 
We label $\mathcal{S}_c$ with the template $\mathcal{S}_t^{*}$ of smallest RMSD. 
To refine the boundary, we slide the end point $e$ forward and backward within a small neighborhood and take $e^{*} = \arg\min_e \mathrm{RMSD}(\mathcal{S}_c, \mathcal{S}_t^{*})$, the location that best fits the chosen type. This $e^{*}$ closes the fitted stitch $\mathcal{S}_i$ and becomes the start anchor of the next iteration.

\section{Conclusion}

\rev{We presented CT2Yarn, a human-in-the-loop framework that recovers a single continuous yarn path from the micro-CT density volume of a real crocheted object.}
\rev{Because the fiber orientations observed in the volume are not aligned with the yarn direction, we estimate a voxel-wise orientation field with Gabor filtering and aggregate it into yarn-level orientations through an anisotropic mean-shift procedure.}
\rev{A topology-preserving skeletonization then extracts yarn-path fragments as a set of disjoint trees, which a heuristic graph-repair stage merges into long continuous sections by exploiting the single-yarn constraint of crochet.}
\rev{For the connections the automatic pass cannot resolve, a graphical user interface enables interactive correction.}

\rev{We evaluated our method on micro-CT scans of 18 hand-crocheted samples covering three stitch types, primarily through visual comparison, supported by quantitative metrics against manually annotated references and against a synthetic volume with an exact ground truth, and by an ablation study of our design choices.}
The recovered yarn path supports downstream applications including physically based simulation, ply-level rendering, and stitch-pattern extraction, and we release both the micro-CT volumes and the reconstructed centerlines to support future research.

\begin{figure}[t]
  \centering
  \includegraphics[width=\linewidth]{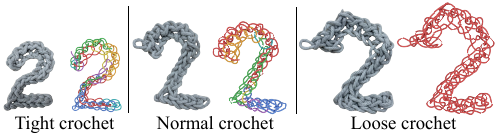}
  \caption{\textbf{Limitation on tightly crocheted samples.}
Reconstruction results for identical crochet patterns made with different stitch tensions: tight (left), normal (middle), and loose (right). Different colors denote different recovered fragments.
Higher stitch tension increases contact between neighboring yarn segments, leading to shorter and more fragmented reconstructions.}
  \label{fig:limitations}
  \vspace*{-0.2cm}
\end{figure}

\paragraph*{Limitations and Future Work}
Our pipeline degrades on tightly crocheted samples.
When stitches are pulled taut, adjacent yarns press together and the CT contrast at their boundary drops to the noise level, so the mean-shift stage loses its separation cue and merges center points that should stay distinct.
As Fig.~\ref{fig:limitations} shows, for the same stitch pattern the way it is crocheted strongly affects the extracted centerline: loose samples are reconstructed without any edits, whereas the curves in tight samples are much shorter and require further manual editing.

\rev{Our stitch-pattern extraction is also limited to approximately planar pieces and a small set of basic stitch types, and extending it to non-planar layouts and a richer stitch vocabulary is left for future work.}
\rev{A further challenge is to generalize the pipeline to fully three-dimensional crocheted objects such as amigurumi.
The single continuous yarn of our samples is not merely an input assumption but the topological constraint behind our loop detection, whereas amigurumi are typically assembled from several separate yarns and stuffed with a fibrous filling that is difficult to distinguish from the yarn in CT.
Reconstructing them would require reasoning about multiple interacting strands and separating yarn from filling.}

\rev{Finally, our semi-automatic workflow produces paired CT volumes and yarn centerlines with little human effort, so it can generate training data at scale.
Such a dataset could support learned centerline detectors for harder, lower-contrast scans, as well as downstream applications such as centerline-aware crochet pattern generation and neural stitch classification.}

\section*{Acknowledgements}
We would like to thank Yumi Endo for her assistance with crafting crochet, as well as the anonymous reviewers for their constructive comments, which helped improve the manuscript. This work was supported by JST SPRING, Grant Number JPMJSP2108.
\printbibliography

\end{document}